\documentclass[wcp]{jmlr}

\makeatletter

\renewcommand{\@oddhead}{}
\renewcommand{\@evenhead}{}
\makeatother

\makeatletter
\let\Ginclude@graphics\@org@Ginclude@graphics
\makeatother

\usepackage{booktabs}
\usepackage{appendix}
\usepackage{amsmath}
\usepackage{natbib}
\usepackage{fancyhdr}
\usepackage{lipsum}
\usepackage{tikz}
\usepackage[load-configurations=version-1]{siunitx} 

\theorembodyfont{\upshape}
\theoremheaderfont{\scshape}
\theorempostheader{:}
\theoremsep{\newline}

 \usepackage{graphicx} 
\newcommand{\comment}[1]{}

\usepackage{xcolor}
 \usepackage{adjustbox}
\usepackage{longtable} 
  
  \usepackage{algorithm}
\usepackage[noend]{algpseudocode}

 \usepackage{amsmath}
 
\usepackage{hhline}
\usepackage{multirow}
\usepackage{caption}
\usepackage[noend]{algpseudocode}

\def\bbbe{\mathbb{E}}
\def\bbbp{\mathbb{P}}

\def\bbbn{\mathbb{N}}
\def\ignore#1{}

\newcommand{\mcL}{\mathcal{L}}

\newcommand{\Loss}{\mathop{\mathrm{Loss}}\nolimits}

\usepackage{array}

\newenvironment{conditions}
  {\par\vspace{\abovedisplayskip}\noindent\begin{tabular}{>{$}l<{$} @{${}-{}$} l}}
  {\end{tabular}\par\vspace{\belowdisplayskip}}

\renewcommand{\binom}[2]{\genfrac{(}{)}{0pt}{}{#1}{#2}}
\newcommand{\stirling}[2]{\genfrac{[}{]}{0pt}{}{#1}{#2}}

\title[Temporal distribution of clusters of investors and their
application in PEA]{Temporal distribution of clusters of investors and their
application in prediction with expert advice}

  \author{\Name{Wojciech Wisniewski} \Email{wojciech.wisniewski.2019@live.rhul.ac.uk}\\
\Name{Yuri Kalnishkan} \Email{yuri.kalnishkan@rhul.ac.uk}\\
  \addr Department of Computer Science \\Royal Holloway, University of London,\\ 
  Egham, United Kingdom 
  \AND
  \Name{David Lindsay} \Email{david@algolabs.com}\\
  \Name{Si\^{a}n Lindsay} \Email{sian@algolabs.com}\\
  \addr AlgoLabs\\Bracknell, United Kingdom}

\begin{document}

\begin{tikzpicture}[remember picture,overlay]
    \node[anchor=north east, inner sep=50pt] at (current page.north east) 
        {\includegraphics[width=0.125\textwidth]{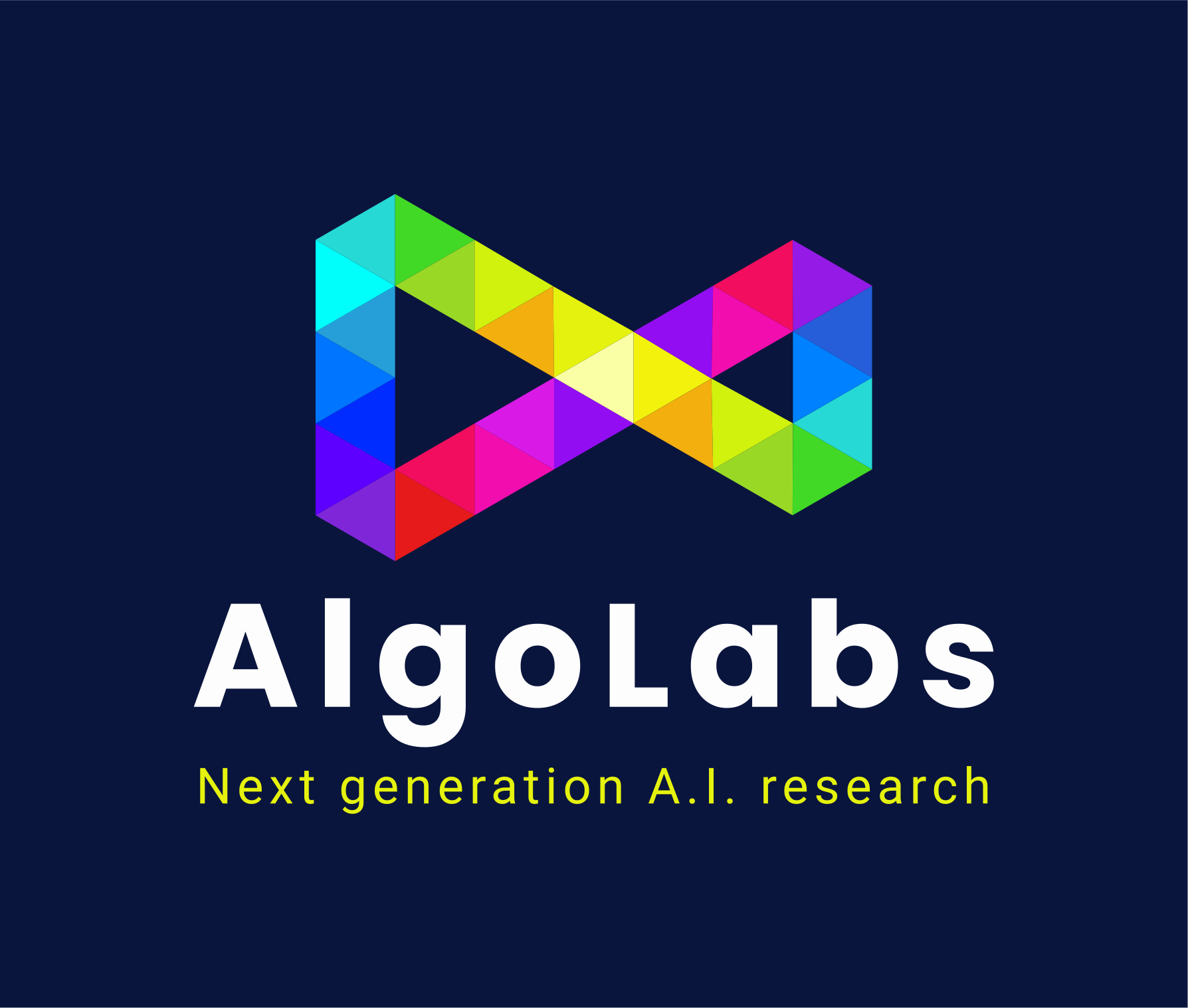}}; 
\end{tikzpicture}

\maketitle

\begin{abstract}
Financial organisations such as brokers face a significant challenge in servicing the investment needs of thousands of their traders worldwide. This task is further compounded since individual traders will have their own risk appetite and investment goals. Traders may look to capture short-term trends in the market which last only seconds to minutes, or they may have longer-term views which last several days to months. To reduce the complexity of this task, client trades can be clustered. By examining such clusters, we would likely observe many traders following common patterns of investment, but how do these patterns vary through time? Knowledge regarding the temporal distributions of such clusters may help financial institutions manage the overall portfolio of risk that accumulates from underlying trader positions. This study contributes to the field by demonstrating that the distribution of clusters derived from the real-world trades of 20k Foreign Exchange (FX) traders (from 2015 to 2017) is described in accordance with Ewens' Sampling Distribution. Further, we show that the Aggregating Algorithm (AA), an on-line prediction with expert advice algorithm, can be applied to the aforementioned real-world data in order to improve the returns of portfolios of trader risk. However we found that the AA 'struggles' when presented with too many trader ``experts'', especially when there are many trades with similar overall patterns. To help overcome this challenge, we have applied and compared the use of Statistically Validated Networks (SVN) with a hierarchical clustering approach on a subset of the data, demonstrating that both approaches can be used to  significantly improve results of the AA in terms of profitability and smoothness of returns.
\end{abstract}

\section{Introduction}
\label{sec:intro}

 In recent years, published research has highlighted a growing interest in methods to cluster together traders based on their strategic and behavioural features, as well as studying how they influence each other.  We summarise the key contributions in this area since 2012.
Since 2012, various studies have investigated clustering of investors in financial markets. \cite{Tumminello2012} used Statistically Validated Networks (SVN) and the Infomap method to detect clusters of investors with similar trading decisions. Similarly, \cite{Bohlin2014} identified clusters by studying the relationship between portfolio and trading decisions of Swedish investors.  \cite{mus2016} used hierarchical clustering and SVN to detect clusters of investors with similar trading decisions. \cite{challet} extended the SVN methodology and inferred a lead-lag network of clusters of investors trading in FX, which showed that most of the trading activity has a market endogenous origin. \cite{Sueshige2018} detected and clustered traders with respect to limit-order and market-order strategies, and classified trading strategies based on their response pattern to historical price changes. \cite{Gutierrez2019} used mutual information and transfer entropy to identify a network of synchronization and anticipation relationships between financial traders. \cite{Baltakiene2019} constructed multilink networks covering 2 years after IPOs and obtained clusters of investors characterized by synchronization in the timing of trading decisions. \cite{cordi2020market} proposed a method to detect lead-lag networks between the states of traders determined at different timescales and observed that institutional and retail traders have different causality structures of lead-lag networks. \cite{Barreau2020} proposed a deep learning architecture, ExNet, for both investor clustering and modeling. Finally, Viet et al. \cite{viet} analyzed the structure of investor networks during a financial crisis and showed changes in investor trading behavior and mutual interactions in the stock market.

In his analysis of the topic entitled ``Clusters of Traders in Financial Markets''  \cite{Mantegna2020}, Mantegna describes a working hypothesis for analysing the dynamics of clusters of investors trading in financial markets, by remarking that the empirical results of \cite{Musciotto2018} are fully consistent
with Aoki's modeling hypothesis in \cite{Aoki2000}. Aoki proposed to use the framework of the Ewens' sampling formula \cite{EWENS197287} for the characterisation of clusters of economic agents having reached a dynamical equilibrium in a specific market. 

Using a proprietary dataset derived from traders investing in the Foreign Exchange (FX) market, we set out to contribute to the body of research concerning the clustering of financial market traders by focusing on the clusters' temporal distributions. For this purpose we constructed sliding window investor networks (Statistically Validated Networks) based on statistically significant trade time synchronisation and showed that the Ewens' Sampling Distribution is a good fit.  Having greater insights into the variations of trader clusters throughout time will likely assist financial institutions as they manage the overall portfolio of risk that builds up from  underlying trader positions. 

Whilst the literature concerning clustering in portfolio selection is very broad, there is no such analysis on how clusters may be leveraged by prediction with expert advice techniques such as the Aggregating Algorithm (AA). This may be because the theoretical dependency of the AA on the number of experts is mild (under uniform initial distribution, the dependency is logarithmic). This suggests that with a larger number of experts, over time the algorithm will manage to work out if they are needed. In the domain of AA in several financial contexts, several researchers have contributed significantly. \cite{VOVK1998}, \cite{vovk1990} have developed the general problem of prediction with expert advice. Specifically, \cite{vovkwatkins} proposed a portfolio selection method using prediction with expert advice, considering realistic trading scenarios. \cite{yugin} constructed a universal trading strategy based on well-calibrated forecasts using prediction with expert advice methods, namely, AdaHedge-type algorithms.  \cite{WAA} considered constant rebalanced portfolios as expert advice and constructed a universal portfolio strategy using the weak aggregating algorithm. Recent papers such as \cite{najim} have demonstrated how the AA can be applied to FX data for improving returns of portfolios of trader risk. However in this practical situation the number of experts present a problem, they overwhelm the advantages of the best expert. Thus we hypothesise that clustering of the trading experts in our dataset should make prediction more reliable and robust by reducing noise. 

The organisation of the paper is as follows. The paper is split into two major parts. First, we demonstrate how the cluster distributions of trading activity can be described according to Ewens' sampling distribution. We also investigate whether these distributions are stationary and depend on the way the clusters are detected. Secondly, we show how clusters of traders over time can be used to improve the profitability of the AA.

\section{Clustering of retail traders by their synchronicity}
\label{sec:2}

We largely follow the methods developed in  \cite{Tumminello2011}, i.e. introduce a behavioural synchronicity measure between traders and then construct a statistically validated network on which an unsupervised clustering method is used.
\subsection{Synchronicity between traders}
A simple way  to infer if  two traders have similar trading behaviours is to compare their scaled trading volume (referred to as the imbalance ratio) in a specific time frame. In order to do so, we partition the  time line  into  disjoint intervals $\cup[t,t+ \delta t[$ and let $r(i,t)$ be the imbalance ratio of trader i in  interval $[t,t+ \delta t[$:
\begin{equation}
  r(i,t)=\frac{b(i,t)-s(i,t)}{b(i,t)+s(i,t)} .
\end{equation}
 where $b(.,.)$ and $s(.,.)$ denote the total volumes bought and sold by the trader in a given time frame.

 For a given threshold $a$ we define a trader state as follows:
 \begin{equation}
    state(i,t)=
    \begin{cases}
       \text{buying state}, & \text{if}\ r(i,t) > a \\
       \text{selling state}, & \text{if}\ r(i,t) <  $- a$ \\
         \text{neutral state}, &  \text{if}\ $- a$ \le r(i,t) \le a \\
           \text{inactive state}, & \text{if}\ b(i,t)+s(i,t)=0
    \end{cases}
  \end{equation}

The synchronicity of a pair of traders is measured by counting the co-occurrences
in the time series of their states, and attributing a $p$-value that reflects the statistical significance of this synchronicity assuming pure randomness. The hypergeometric distribution is used to calculate the $p$-value. The $p$-value is the probability that in a series of $n$ trades, where one trader was $n_p$ times in a state $p$ and the other was $n_q$ times in a state $q$, these occurrences overlapped $n_{p,q}$ times or more: 
$$p(n_{p,q})=1-\sum_{i=0}^{n_{p,q}-1} H(i|n,n_p,n_q)$$
where
$$H(i|n,n_p,n_q)=\frac{\binom{n_p}{i} \binom{n-n_p}{n_q-i}}{\binom{n}{n_q}}$$
This method is often used to measure similarity in genetic sequences. In literature there exist many other alternative methods; however, the hyper-geometric test can be used with sparse data and it is not sensitive to outliers. Therefore it suits our needs.

To deal with the testing of all pairs of traders and all types of co-occurrences a multiple hypothesis testing correction is needed. For this purpose we use the Bonferroni correction which is the statistical significance (0.05) divided by the number of tests. It is worth pointing out alternatively one could use the false discovery rate for multiple test correction. 
 
\subsection{Statistically validated  network}
A statistically validated  network is a network built by validating links between pairs of traders if the $p$-value of their synchronisation is smaller than the corrected threshold. Traders without any links are dropped. In the resulting network we exclude links between opposite actions (buy-sell), links between neutral states and links between inactive states. The reason being that we are mostly interested in active traders with the same kind of behaviour (buy-buy and sell-sell) and which manifest high trading activity.


\section{Clustering distribution}

This section will define Ewens' sampling formula, which is  the backbone of Aoki's modeling hypothesis concerning the dynamics of trading behaviour.

\subsection{Partition vector}

We will now introduce a partition vector which will be useful for our modelling purposes. 

Let $c_i$ be the number of clusters with exactly $i$ traders. Then $K_n=\sum_{i=1}^n c_i$ is the total number of clusters formed by $n$ traders and $\sum_{i=1}^n i  c_i=n$. We will call the vector
$c=( c_1 , c_2 ,\ldots, c_n )$ a partition vector.

\subsection{The Ewens Sampling distribution} 
Ewens' sampling formula describes a specific probability for the partition of the positive integer into parts. It was discovered by Ewens \cite{EWENS197287} as providing the probability of the partition of a sample of $n$ selectively equivalent genes into a number of different gene types (alleles). For positive integers $c_1, c_2, . . .,c_n$ , satisfying $\sum_j jc_k=n$ and being a realisation of a random partition vector $({C}_1(n),{C}_2(n),\ldots,{C}_n(n))$, we have:
\begin{equation}\label{esflaw}
    \bbbp_\theta({C}_1(n)=c_1,\ldots,{C}_n(n)=c_n) = \frac{n!}{\theta_{(n)}}\,\prod_{j=1}^n \left(\frac{\theta}{j}\right)^{c_j} \frac{1}{c_j!},
\end{equation}
for $\theta \in (0,\infty)$, $\theta_{(n)} := \theta(\theta+1)\cdots(\theta+n - 1) = \Gamma(n+\theta)/\Gamma(\theta), n \geq 1$ and $\theta_{(0)} = 1$.\footnote{We define $\theta_{(-k)} = 0$, for $k \in \bbbn$.}  

We  are interested in groups of traders with strategies manifesting similar synchronisation, and since the SVN clustering process builds a network with strong links between pairs of traders we conjecture there should be no (or only a small number of)  mono-communities.
Thus it is natural to fit a distribution conditional on the event: $c_1=0$.

Let us define the conditional distribution   $(\tilde{C}_2(n),\ldots,\tilde{C}_n(n))$  (see \cite{dasilva2020random}) in the following  way:
\begin{equation}\label{derlaw}
\mcL(\tilde{C}_2(n),\ldots,\tilde{C}_n(n)) = \mcL(C_2(n),\ldots,C_n(n) \vert C_1(n) = 0)
\end{equation}
The probability of the condition is given by:
\begin{equation}
    \lambda_n(\theta) := \bbbp(C_1(n) = 0) = \frac{1}{\theta_{(n)}} \sum_{k = 1}^n \theta^k D(n,k),
\end{equation}
(with $\lambda_0(\theta) = 1$ and  $\lambda_1(\theta) = 0.$), where $D(n,k)$ is the number of derangements of size $n$ having $k$ cycles:
\begin{equation}
    D(n,k) :=  \sum_{l=0}^k (-1)^l \binom{n}{l} \stirling{n-l}{k-l}\enspace;
\end{equation}
here $\stirling{n}{k}$ is the unsigned Stirling number of the first kind.

Accurate computation of alternating series, present in $D(n,k)$, is a well-known hard problem therefore it is useful to give the recursive relation of $ \lambda_n(\theta)$ which verifies:
\begin{equation}
    \lambda_{n+1}(\theta) := \frac{n}{n+\theta} \Bigg[ \lambda_n(\theta) +\frac{\theta}{n+\theta-1} \lambda_{n-1}(\theta) \Bigg].
\end{equation}

Table \ref{tab:itemize} summarises relevant characteristics of the Ewens' conditional and non conditional distribution.

    \begin{table}[htbp]
\floatconts
{tab:itemize}  
{\caption{Comparison of  Ewens conditional and non-conditional distributions}}
{\begin{tabular}{ccc} \toprule
    {$Feature$} & {\textit{Ewens distribution}} &   {\textit{ Conditional Ewens distribution }} \\ \midrule
    Probability  & $
     \frac{n!}{\theta_{(n)}} \prod_{j=1}^n \left(\frac{\theta}{j}\right)^{c_j} \frac{1}{a_j!},$
  &   
     $\frac{n!}{\theta_{(n)}\lambda_n(\theta)}\prod_{j=2}^n \left(\frac{\theta}{j}\right)^{c_j} \frac{1}{a_j!}.$
 \\ \midrule
  
    Expected  cycle counts  & 
$\bbbe C_j(n)=\frac{n!}{ \theta_{(n)}}\frac{ \theta_{(n-j)}}{(n-j)!}\, \frac{\theta}{j},$
  &  $
\frac{\lambda_{(n-j)}(\theta)}{\lambda_{(n)}(\theta)}\bbbe C_j(n). $  \\    \midrule
    Expected number of cycles &  $\bbbe K_n=
\sum_{i=0}^{n-1} \frac{{\theta}}{{\theta}+i},$
  &  
$   \sum_{j=2}^n \frac{\lambda_{(n-j)}(\theta)}{\lambda_{(n)}(\theta)}  \bbbe (C_j(n) ).$
  \\ \bottomrule
\end{tabular}}
\end{table}

\section{Experiments}
In this section we describe the proprietary dataset and the experiments.

\begin{figure}[htbp]
  \floatconts
    {fig:both5}
    { }
    {\begin{minipage}[b]{0.4\linewidth}
      \centering
      \includegraphics[width=\linewidth]{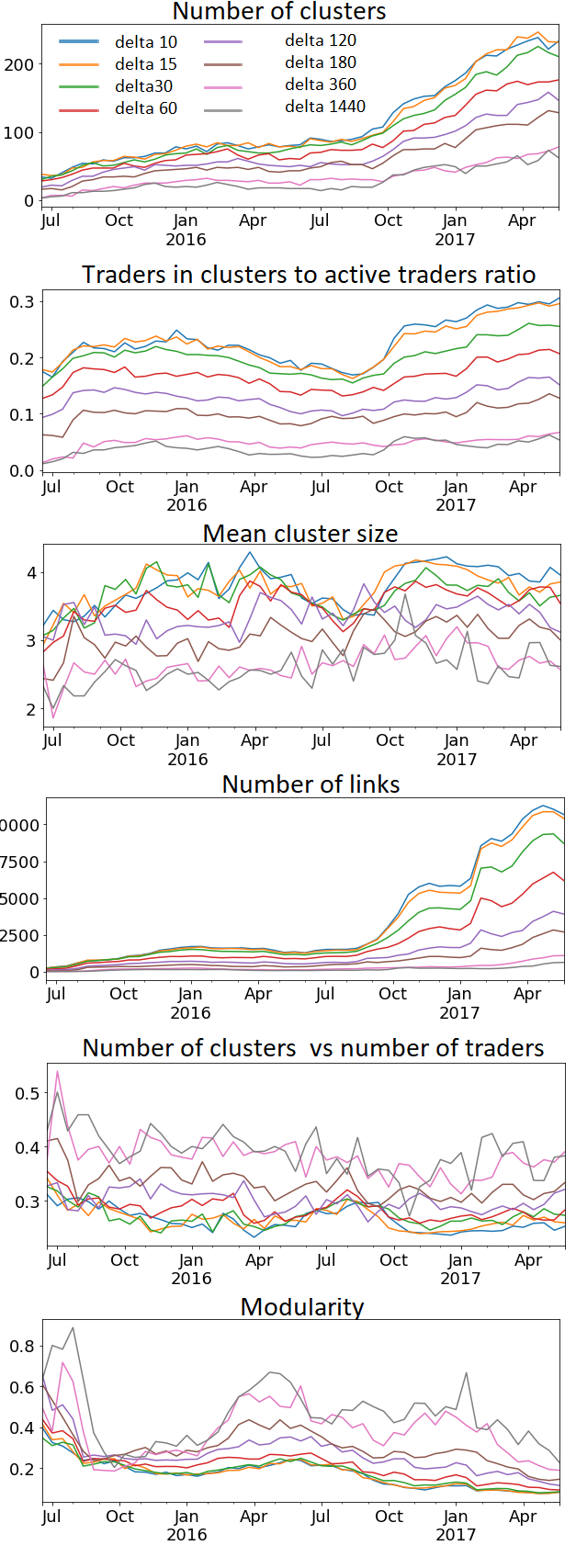}
      \captionsetup{justification=centering}
      \caption{ Evolution of some statistics (number of clusters, number of links, number of traders in clusters vs active traders ratio, number of clusters vs number of traders, mean cluster size, modularity) over time for a network of traders at deltas (10, 15, 30, 60, 120, 180, 360 and 1440 minutes) for EUR/USD currency pair.  }
      \label{fig:statssvn}
    \end{minipage}
    \hfill
    \begin{minipage}[b]{0.55\linewidth}
      \centering
       \includegraphics[width=\linewidth]{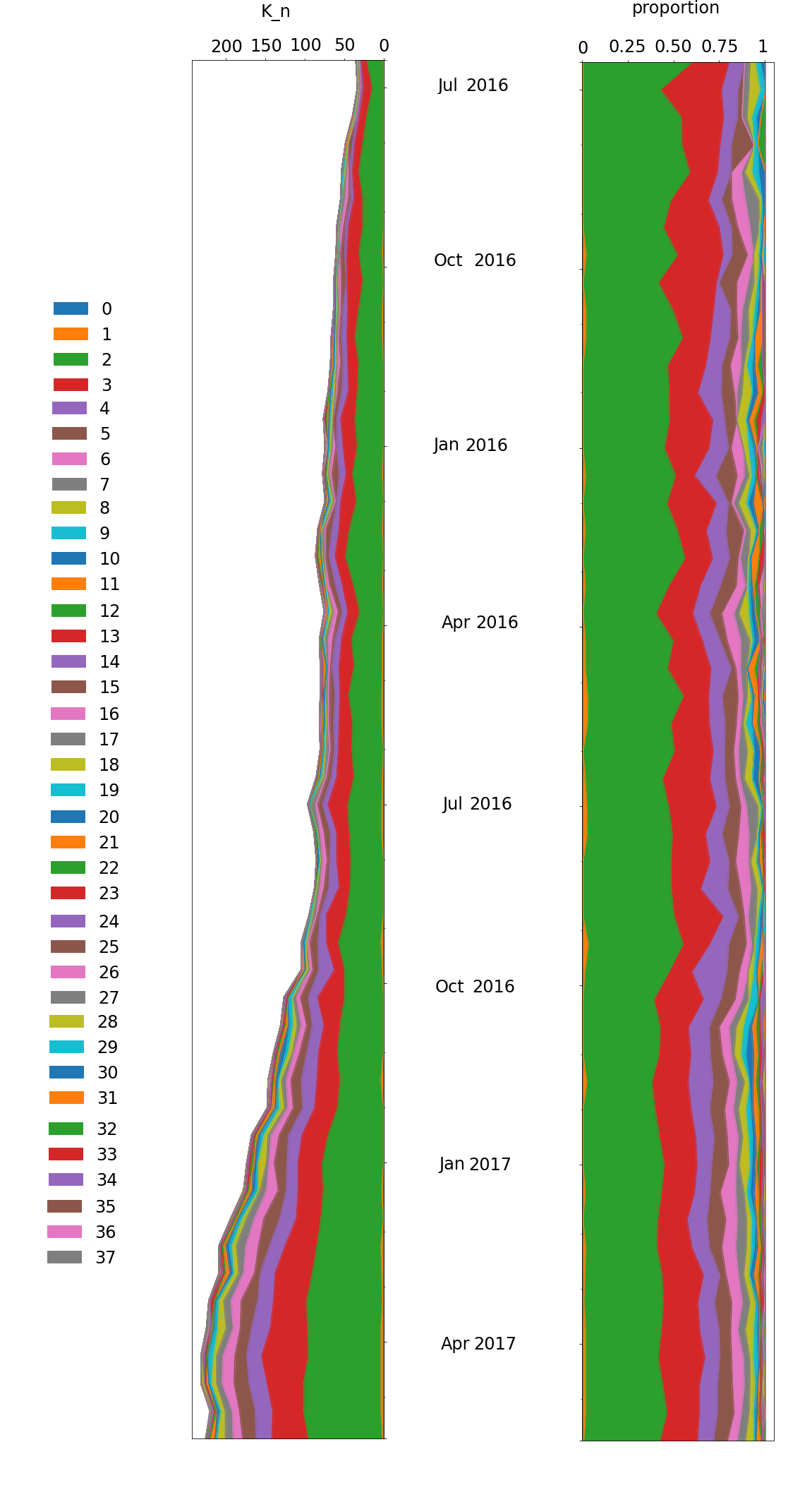}
      \captionsetup{justification=centering}
      \caption{ Proportion vector and normalised proportion vector of temporal evolution for clustering on EURUSD  for 10min delta and cutoff 100. The Infomap algorithm was used to identify clusters after the SVN networks was constructed. }
      \label{fig:propv1}
    \end{minipage}}
\end{figure}

\subsection{The dataset description  }

We consider financial data gathered from the client trades of a retail Foreign Exchange (FX) broker. A typical retail broker will provide their clients with an online trading platform software such as MetaTrader 4 (MT4) where they can place trades, monitor positions, track both historic and live movements in prices, and access the latest world economic news. Online trading platforms often operate under the stipulation that once an order is placed (opened), it must be closed in its entirety. Source data is essentially stored in a temporal table with each row representing a client order that provides the opening and closing time, as well as the currency traded (symbol), amount traded and side (buy or sell) of the order.

The proprietary dataset comprises the trades made by over 20k clients during 2015-2017. Each  client  was  allowed to buy or sell  any  of  available currency  pairs and they could  place  trades  as  many  times  as  they  wanted,  at any time  of  day  provided  they  stayed  within  the  confines  of their  leveraged  funds. The dataset contains only necessary features for further investigation  namely an investor's anonymised ID, opening and closing trade times, amount of lots traded, sign (long or short position), and the traded symbol.

\subsection{Experimental protocol}
It is convenient to use sliding windows in order to track the temporal evolution of clustering. For each in-sample time window, we filtered out traders with less than 100, 500 or 1000 trades (referred to as the cut-off). We observe that the number of traders grows in an approximately linear fashion throughout time which is related directly with the business growth. We focus our investigation on trading activity that occurs during  standard business days within the most active hours (6am - 6pm). Investigations are conducted solely considering the EUR/USD currency pair. We construct a sliding window of size 6 months and shift it every 2 weeks. Then we build a SVN network at every step using the imbalance ratio time series for $\delta t$ ranging from 10, 15, 30, 60, 120, 180, 360, and 1440 minutes (referred to as deltas).

\subsection{SVN clustering and its descriptive statistics over time}

To categorize traders into distinct groups, we used Infomap clustering algorithm  \cite{Rosvall2008} since its  popularity can be attributed to its information-theoretic approach, scalability, high quality clusters, flexibility, and statistical significance.  According to the study \cite{comparaisonclusteringakgos} the Infomap clustering algorithm empirically gave the best results in \cite{comparaisonclusteringakgos} when applied to different benchmarks on Community Detection methods. Our empirical findings indicate that  evolution of the proportion vector (with respect to its normalised version) allure satisfies our conjecture of a sparse number of  mono-communities (see Figure \ref{fig:propv1}).  From the figures, we notice a smooth evolution of proportions, and also the appearance of new and larger clusters - this is to be expected since the number of traders is growing over time. Moreover we observe a pattern of having less clusters of significant cardinality. An existence of a very big cluster (and many very small ones)  would negate the heterogeneity of trading strategies. We observe  that Infomap is consistent with the resolution scale and number of trades cut-off.  We calculated several pertinent statistics to evaluate how the SVN's are affected by different time resolutions sampled throughout the lifespan of the entire dataset (i.e. from 2015 - 2017), as illustrated in Figure \ref{fig:statssvn}.
As previously stated, the number of traders in the dataset increases over time  however we notice a sudden increase in the number of links and clusters from July 2016.

This results in an increase in the number of clusters and links in the sliding networks. We remark stability over time in the ratio of numbers of traders against the number of clusters. At each slide an SVN is built and some traders are never taken into consideration and the ratio of existent traders is increasing slightly with increase of the resolution delta. The modularity  is slowly decreasing and is low  besides deltas of 360 and 1440 minutes, which testifies about rather weak connections between clusters.

\subsection{Goodness of fit }
In order to assess the goodness of fit to the data  we refer to what  is conventionally used: a classical  $\chi^2$ test. The parameter $\theta$ was estimated for every sliding window and since the formula is not explicit  for $\bbbe  K_n$ (see table \ref{tab:itemize}) we approximate it to the closest integer. It is worth noting that for a non conditional Ewens distribution one can readily find an explicit formula for $\theta$ using $\bbbe  K_n$.

Taking the example  for   $\delta$ equal to 10 mins and cut-off of minimum 100 trades we apply the  $\chi^2$ test for 50 sliding windows at  significance of  0.05. We find a $95\%$ pass rate which  confirms that  most of the time the conditional Ewens distribution is a good fit.

\begin{figure}[htbp]
  \floatconts
    {fig:both2}
    { }
    {\begin{minipage}[b]{0.6\linewidth}
      \centering
      \includegraphics[width=\linewidth]{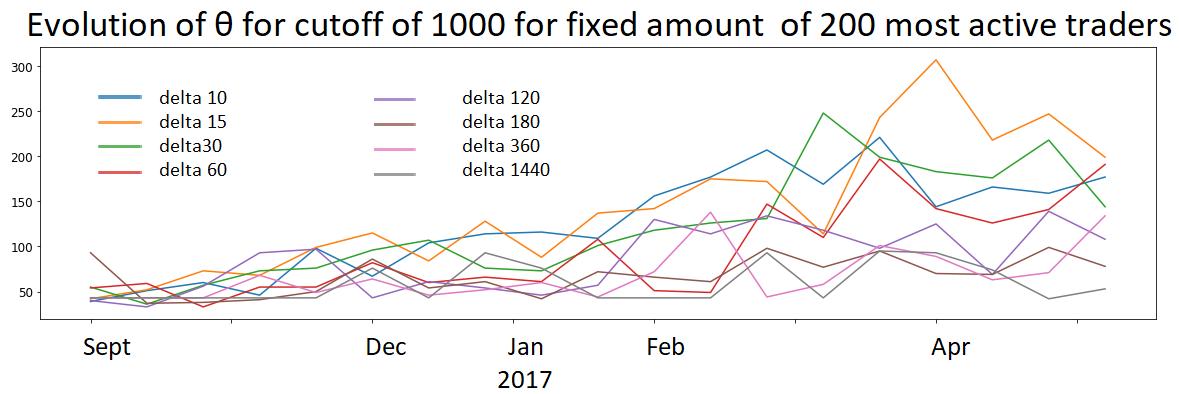}
      \captionsetup{justification=centering}
      \caption{Evolution of $\theta$ parameter for all $\delta $ time slices and cutoff of 1000  in the fixed amount of 200 most active traders. Other scenarios bear similarities in the shape of the curves i.e. for deltas 360 and 1440  the parameter $\theta$  stays more or less stationary and others increase suddenly at some point.  }
     \label{fig:thetas}    
    \end{minipage}
    \hfill
    \begin{minipage}[b]{0.35\linewidth}
      \centering
       \includegraphics[width=\linewidth]{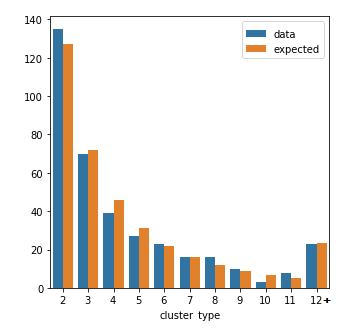}
      \captionsetup{justification=centering}
      \caption{A comparison of empirical and theoretical fit on last sliding window. The plots were obtained using EURUSD  data for 10min scale and cutoff 100. }
      \label{fig:fit}
    \end{minipage}}
\end{figure}

Figure \ref{fig:passrate} shows that for all studied scenarios in most cases we have a high pass-rate. In general  for a cut-off of 100 the pass-rate is above $85\%$, for others it seems to increase with delta. Figure \ref{fig:fit} illustrates a typical comparison between empirical and theoretical fit on a given sliding window which is satisfying. Figure \ref{fig:thetas} shows the evolution of the Ewens distribution fitted parameter. It is more or less stationary for bigger deltas and  increasing for smaller ones. Larger estimated parameter  $\hat{\theta}$ indicates a higher so-called mutation rate, therefore the existence of more clusters.

\subsection{Temporal cluster evolution and  consistent grouping identification issue}\label{groupingidissue}
In some cases we require consistent grouping identification and the main difficulty comes from the lack of consistent naming of clusters for subsequent time frames. The latter allows us to, amongst other things, produce meaningful visualisations. The technique used relies on a total consistency measure which is in close relation to the Jaccard index (for more details see \cite{liechti2020time}).

\begin{figure}[htbp]
  \floatconts
    {fig:both}
    { }
    {\begin{minipage}[b]{0.6\linewidth}
      \centering
      \includegraphics[width=\linewidth]{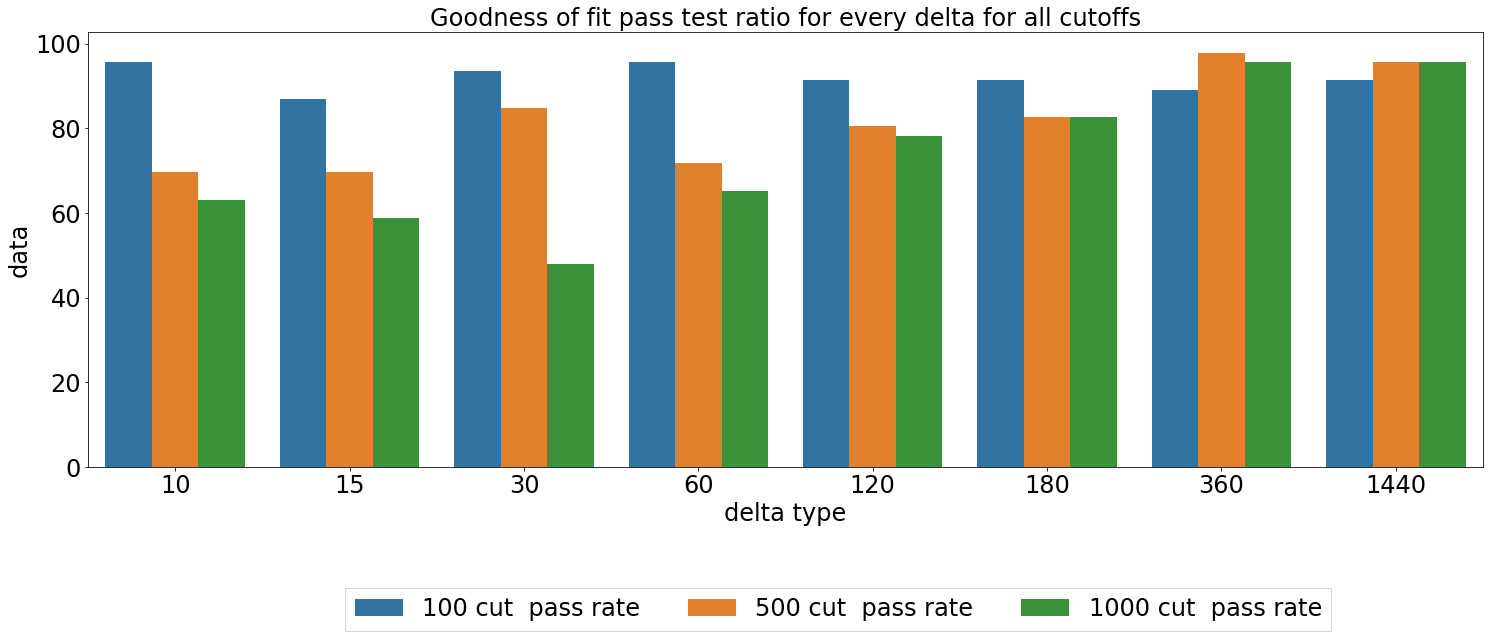}
      \captionsetup{justification=centering}
      \caption{Pass rate in percent for all $\delta $ time slices and 100, 500 and 1000 cutoffs. This rate represents the ratio of non rejected null $\chi^2$ hypothesis for all sliding windows }
      \label{fig:passrate}
    \end{minipage}
    \hfill
    \begin{minipage}[b]{0.35\linewidth}
      \centering
       \includegraphics[width=\linewidth]{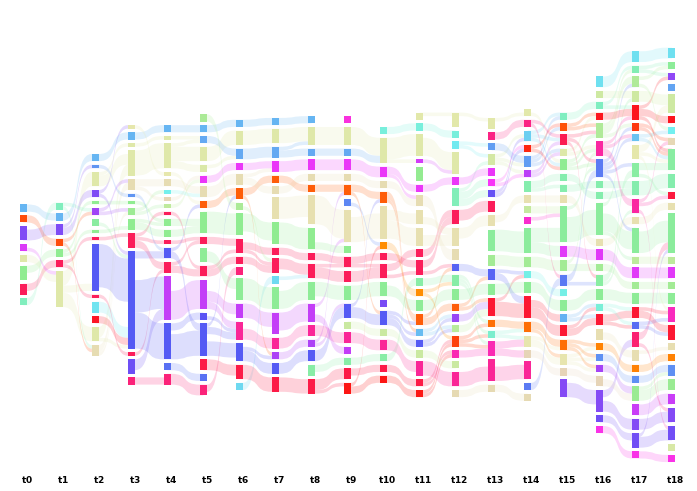}
      \captionsetup{justification=centering}
      \caption{Alluvial plot with 1 step history (see \cite{liechti2020time} for  more details) for 200 most active traders with cutoff 1000 and delta of one day. }
      \label{fig:alluv}
    \end{minipage}}
\end{figure}

In Figure \ref{fig:alluv} we see a so-called alluvial plot where at a given time, traders belonging to the same group are stacked together to form a continuous flow. The stability of group composition is shown when the same colouring persists between two time steps. However a group can split, merge, die out, appear suddenly or persist throughout time. These changes in groups are to be expected as traders' investment strategies evolve over time, and existing traders leave and new traders join. 
Overall we remark some stability, however as expected eventually there are die outs, merges, splits and new appearances. 
When we considered different deltas (results not shown), we found that  larger groups were more prevalent for smaller time frames.

\section{Clusterised Aggregating Algorithm}

We wish to study the temporal evolution of clusters of trading activity and investigate how they can be used for practical purposes. Clustering evolution could be used in prediction problems since grouping has the advantage of simplifying the description of the system state by reducing the dimensionality of the prediction problem. In the literature there are numerous examples of the latter set in a financial context. For example, in \cite{challet} the authors used SVN's to demonstrate improvement in predicting both the sign of the order flow and the direction of the average transaction price for a retail trader dataset. In this study we have applied the clustering evolution to prediction with an online expert advice model, namely the Aggregating Algorithm (AA) \cite{vovk1990} and \cite{VOVK1998}. 
The AA is given a series of online predictions from a pool of experts (in our case the traders). At each time epoch, the loss of each experts' prediction (in our case a trader's investment decision) is fed back into the AA and over time adjusts its trust in each expert to make future predictions.  In the next subsections we introduce the framework of the AA and the games of investment with expert advice.

\subsection{Aggregating Algorithm}

Suppose  that  the learner $L$ is  tasked  with  predicting  elements  of  a  sequence  $\omega_1,\omega_2,\ldots$ called outcomes.  The outcomes occur in discrete time.  Before seeing outcome $\omega_t$,  the learner is outputting a prediction $\gamma_t$.  The quality of the prediction is measured by a loss function $\lambda(.,.)$.  The expert aims to suffer low cumulative loss:
$$\Loss_T(L)=\sum_{t=1}^T\lambda(\omega_t,\gamma_t)$$

We assume that the set of all possible outcomes (outcome space) $\Omega$ is known to us in advance and we are allowed to draw predictions from a known prediction space $\Gamma$,  which  may  or  may  not  be  the  same  as   $\Omega$.   The  function $\lambda$ is  also known and maps $\Gamma \times \Omega$ to a subset of the extended real line, typically $[0,+\infty]$. The choice of a triple $G=\langle\Omega,\Gamma,\lambda\rangle$, is referred to as a game.

Suppose  that  the  learner  gets  help  from experts.   The  experts  predict  the same sequence and their predictions are made available to the learner before it commits to its own predictions.   We are not concerned with their internal mechanics, which may well be inaccessible to us (e.g., the experts may rely on some sources of information unavailable or even unknown to us).  The interaction with experts may be described by the following protocol.  Here we assume that experts are parameterised by  $\theta \in \Theta$.

\comment{
\begin{algorithm2e}[t]
\caption{Prediction with Expert Advice Protocol}
\label{alg:AA_protocol}
\DontPrintSemicolon
\LinesNumbered
\For{$t=1,2,\dots$}{
    Experts $E_{\theta},\theta \in \Theta$, announce predictions $\gamma_t^{\theta} \in \Gamma $\;
    Learner outputs $\gamma_t \in \Gamma $\;
    Nature announces $\omega_t \in \Omega$\;
    For each Expert $E_{\theta}$ suffers loss $\lambda(\gamma_t^{\theta},\omega_t)$\;
    Learner suffers loss $\lambda(\gamma_t,\omega_t)$\;
}
\end{algorithm2e}
}

Expert $E_{\theta}$   suffers loss $\Loss_T(E_{\theta}) =\sum_{t=1}^T \lambda(\gamma^{\theta}_t,\omega_t)$.  The goal of the learner is to merge experts' predictions $\gamma^{\theta}_t$ into its own prediction $\gamma_t$   in such a way that the learner's loss $\Loss_T(L)$ is low as compared to retrospectively best experts. It may use information about past outcomes and predictions. Formally, we are seeking a merging strategy: 
$$ S: (\Gamma^\Theta \times \Omega)^* \times \Gamma^\Theta   \rightarrow \Gamma$$

We  typically  want S to  guarantee  an  upper  bound  on  $\Loss_T(L)$  in  terms  of $\inf\limits_{\underset{\theta \in \Theta}{}} \Loss_T(E_\theta)$;  we want $\Loss_T(L)$ to be low whenever $\Loss_T(E_\theta)$ is low for some $\theta$. We assume that the pool of experts is finite, i.e., $|\Theta|= n <+\infty$.

Consider a game $G=\langle\Omega,\Gamma,\lambda\rangle$ a constant $C >0$ is admissible for a learning rate $\eta >0$  if for every $N= 1,2,\ldots,$ every set of  predictions $\gamma_1,\ldots,\gamma_n \in \Gamma$,  and every  distribution  $(p_1,p_2,\ldots,p_n) \in \Delta_{n-1}$, there is $\gamma \in \Gamma$ ensuring for all outcomes $\omega \in \Omega$ the inequality:
$$\lambda(\gamma,\omega) \leq \frac{C}{\eta} \ln\sum_{i=1}^N p_i e^{-\eta\lambda(\gamma,\omega)}$$
The mixability constant  $C_\eta$ is the infimum of all $C >0$ admissible for $\eta$.  This infimum is usually achieved. The admissibility is required to ensure the learner's predictions exist and belong to $\Gamma$ since for example the learner's prediction of the form  $\gamma_t=\sum_{i=1}^N p_i\gamma^i_t$ is a linear combination and $\Gamma$ may not be convex. 
 The AA takes as parameters a set of prior experts' weights $(q_1,\ldots,q_N) \in \Delta_{N-1}$, a learning rate $\eta >0$ and an admissible $C >0$. The algorithm works as shown in the pseudocode below.

\begin{algorithm2e}[t]
\caption{Aggregating Algorithm}
\label{alg:AA_def}
\DontPrintSemicolon
\LinesNumbered
\KwIn{$\eta,C,q,N$}
initialization of weights $\omega_0^i \sim q_i$ for $i=1,\ldots,N$\;
choice of loss $\lambda(.,.)$\;
\For{$t=1,2,\dots$}{
   read experts' predictions $\gamma_t^i$\;
normalise the weights $p_t^i=\frac{\omega_{t-1}^i}{\sum_{j }\omega_{t-1}^j}$\;
output $\gamma_t \in \Gamma$ satisfying for all $\omega\in\Omega$\:
$\lambda(\gamma,\omega) \leq \frac{C}{\eta} \ln\sum_{i=1}^N p_i e^{-\eta\lambda(\gamma,\omega)}$\;
observe outcome $\omega_t$\;
update the weights $\omega_{t}^i=\omega_{t-1}^i\cdot e^{-\eta\cdot \lambda(\gamma_t^i,\omega_t)}$ \;
}
\end{algorithm2e}

The validity of the AA holds under some mild regularity assumptions on the game and  assuming the uniform initial distribution, it can be shown (as in Equation \ref{eq:special}) that the constants in the following inequality are optimal: 
 
\begin{equation} \label{eq:special}
  \Loss_T(L) \leq C\Loss_T(E_i)+\frac{C}{\eta}\ln N  
\end{equation}

\subsection{Long Short Game}
 
The problem of portfolio selection is a natural special case of a prediction with expert advice problem where in \cite{vovkwatkins} considered realistic trading  scenarios i.e. the Long Short game.

The Long-Short game aims to represent a realistic trading scenario. A trader is allowed to open positions, both long and short, within certain limits based on their deposit and money they had earned previously. The limits aim to minimise the chance of bankruptcy. Given the wealth $W_{t-1}$ at time $t-1$ trader $i$ opens a position  of size $W_{t-1}\gamma^i_t$ when the return $\omega_t$ is known, the trader's wealth changes accordingly:
$$W_t=W_{t-1}\cdot\lambda(\gamma_t^i,\omega_t)=W_{t-1}\cdot(1 + \gamma_t\cdot \omega_t) $$

In this framework one can apply the AA with $\eta=1, C=1$ and
the substitution rule given by  $\gamma_t=\sum_{i=1}^N p_i\gamma^i_t$ to the general long-short game. 
If $1 +  \gamma_t\cdot \omega_t > 0$ for $t=1,\ldots,T$ i.e., the learner does not get bankrupt along the way, the bound \eqref{eq:special} will hold.

  \subsection{AA with Sleeping Experts}

In \cite{najim}, an evaluation of the performance of the AA was made using a real-life trading dataset. Some modifications of the AA were proposed in order to improve the  practical performance of the resulting portfolio. In particular, a downside loss and weighted average between the latter and the long short loss were introduced.
Downside loss, in contrast to long short loss (originally used in \cite{vovkwatkins}), penalises financial losses but does not reward gains since a strategy not to lose money may be more important than the ability to earn money.
 
\begin{equation}
\begin{aligned}
 \lambda_\mathrm{Long~Short~Loss}(\rho,\gamma,r)&=-\log[\max(1+\rho \cdot \gamma \cdot r,0)]   \\
  \lambda_\mathrm{Downside~Loss}(\rho,\gamma,r)&=-\log\{\max[1+\rho \cdot \min(\gamma \cdot r,0),0]\}   \label{loss2}
  \end{aligned}
\end{equation}
where:
\begin{conditions}
\rho     &  scaling factor \\
\gamma     &  investment decision $\in$ [-1, 1] \\   
r &  return
\end{conditions}

In our research we faced one particular challenge with our dataset: the pool of traders constantly changes through time. For example, traders may choose to cease trading with the broker at any time, they may take breaks from trading, new ones may join, or traders may close their account entirely. The AA requires such experts to continually provide predictions through time - a natural way to encode such activities is to use the so-called ``sleeping'' experts extension.

\begin{algorithm2e}[t]
\caption{Aggregating Algorithm With Sleeping Experts}
\label{alg:AA}
\DontPrintSemicolon
\LinesNumbered
\KwIn{$\eta,\rho,n$} 
Initialization of weigths $\omega_0^i=1$ for $i=1,\ldots,n$\;
 Choice of loss $\lambda(\gamma,r)$\;
\For{$t=1,2,\dots$}{
   Get set of awake experts $A_t$ and sleeping experts $S_t$\;
 Get set of awake experts $A_t$ and sleeping experts $S_t$\;
 Read investment of awake experts $\gamma_t^i$ for $i \in A_t$ \label{investement}\;
  Normalise the weights of awake experts $p_t^i=\frac{\omega_{t-1}^i}{\sum_{j:A_t  }\omega_{t-1}^j}$\;
  Calculate investment prediction $\gamma_t={\sum_{j:A_t}p_t^j \cdot  \gamma_{t-1}^j}$\;
  Observe return $r_t$\;
 Update for $i \in A_t$ the weights    $\omega_{t}^i=\omega_{t-1}^i\cdot \exp[-\eta\cdot \lambda(\gamma_t^i,r_t)]$\;
 Update for $i \in S_t$ the weights    $\omega_{t}^i=\omega_{t-1}^i\cdot \exp[-\eta\cdot\lambda(\gamma_t,r_t)]$\;
}
\end{algorithm2e}

\subsection{Clusterised Aggregating Algorithm (CAA) and decision rules \label{decisions}}

The classical AA learner prediction   is:
\begin{eqnarray}
{\gamma_t} = \sum_k{{p^k_t\gamma}^k_{t-1} }
 \end{eqnarray}
 Which is a weighted average of experts' predictions. For clusterised aggregating algorithm (CAA) we introduced two decision rules:

\begin{align*}
    {\gamma_t}^{\mathrm{MEAN}} &= \sum_i \sum_j^{n_i}p^{i,j}_t\cdot \sum_k^{n_i} \frac{{\gamma}^{i,k}_{t-1} }{n_i} && \text{take the mean of experts' predictions in a cluster}\\
    {\gamma_t}^{\mathrm{PEN}} &= \sum_i \sum_j^{n_i}p^{i,j}_t  \frac{{\gamma}^{i,j}_{t-1} }{n_i} && \text{penalise by dividing by the cardinality of a cluster}
\end{align*}
where $n_i$ is the cardinality of $i$-th  cluster and $p^i$ is the sum of probabilities of $i$-th cluster.

The decision rule of ${\gamma_t}^{\mathrm{MEAN}}$ is interesting in a trivial case scenario i.e. having the same duplicated experts in every cluster.
Let's suppose that we have $m$ identical experts in the pool. It appears desirable to collate them into one. However, this is done by the AA automatically. The behaviour of the AA would be the same as if one expert with the combined weight is present in
the pool. Assuming the uniform distribution on the initial experts, the weight
of the combined expert will be $m/N$ and the loss bound for the duplicated experts
$E_i$ (again assuming the mixable case $C = 1$) turns into:
$$\Loss_T(L) \leq \Loss_T(E_i)+\frac{1}{\eta}\ln\frac{N}{m}$$
However, if duplicate experts are bad, this creates a problem: needlessly
increasing $n$ worsens the bound for good experts. For example, if there were two clusters, with each having different duplicated experts and the bigger cluster had better-performing experts then the AA bound would be improved.

The second decision rule i.e. ${\gamma_t}^{\mathrm{PEN}}$  has an interpretation of  partially awake experts if the penalising factor is normalised i.e.
$ \frac{\frac{1}{n_i}}{\sum_{k \in\mathrm{Clusters }}\frac{1}{n_k}} $. This idea was generalised in  \cite{almostsleeping}.  Apart from a prediction $\gamma_t$ such an expert produces a confidence value $ c_t \in [0, 1] $, which quantifies its confidence (a fully sleeping expert would output confidence of 0 and a fully awake expert would output a confidence of 1). Here the confidence would be inverse proportional to the cardinality of the cluster. This is similar to inverse-variance weighting in portfolio selection problems in particular the equal risk contributions portfolio \cite{Maillard60}.

\subsection{Experts as Clusters approach to AA (ECAA)}
Up until now we only clusterised via the decision rules, and the experts were identified as the traders. It seems natural to consider treating clusters of traders as meta-experts. We averaged experts' investement decisions per cluster in order to obtain the meta-experts' predictions.   In appendix \ref{eq:conditionAA}, we derive a condition to which these extensions to the AA would outperform the original set up of the AA with duplicated experts. In practice  we identified the flow of meta-experts according to the alluvial plot (see Figure \ref{fig:alluv}). There are several things to consider in this scenario especially the splitting and merging of clusters on every epoch. We suggested the following approach: 
\begin{itemize}
    \item If the cluster is split then the children would inherit the parents weight divided by number of splits. 
    \item If clusters are merged then the resultant weight is the sum of the parents weights.
\end{itemize}

\subsection{Experiments}

First we applied a data staging technique known as DAPRA (see \cite{dapra}) which, when applied to data streams pertaining to trades and prices, allows one to sample the data at regular time intervals (required for this study). We then compared the performance of the AA with its clusterised counterparts (CAA and ECAA) with the expectation that these extensions would improve scalability and reduce noise. The CAA extension simply takes the mean of investments of awake experts $\gamma$ in a given cluster (MEAN), or divides their decision by the cardinality of the cluster (PEN). As a benchmark we used the equally weighted portfolio strategy. 
We compared the CAA and the ECAA using the SVN-infomap approach with hierarchical clustering based on correlations of the traders' net positions (i.e. difference between total open long (buy) and open short (sell) positions in USD dollars) with a chosen distance metric: $1-|\mathrm{correlation}|$. The latter approach has a possibility of adjusting the construction of clusters by changing the dissimilarity threshold. The rationale behind clustering based on net position correlation is that it is a desirable feature for the broker since it is a measure of risk. The SVN approach is focused on trading synchronicity therefore we have less control on the quality of clustering in regards to the net position. Ideally all traders would trade all the time or have a high trading intersection period but since it is not the case one can end up with ``noisy'' clusters.

\begin{table}[htbp]
\floatconts
{Tbl:summary}
{\caption{Table summarising the experimental results  for CAA.} }
{ 
\begin{adjustbox}{width=0.95\textwidth}
\begin{tabular}{llrrrrl}
\toprule
Strategy & Type & Scaling factor & Return & Sharpe Ratio & Max Drawdown & Calmar Ratio \\
\midrule
EW & Benchmark & - & 1.4\% & 0.6 & 1.2\% & 1.2 \\
AA & Sleeping Experts & 70& \textbf{2.8}\% & 1.1 & 1.8\% & 1.5 \\
CAA & MEAN/ SVN & 70 & 3\% & 1.2 & 1.85\% & 1.8 \\
CAA & MEAN/Hierarchical & 70 & \textbf{4.8}\% & \textbf{2} & 1.15\% &\textbf{4} \\
CAA & PEN/SVN & 70 & \textbf{2.5}\% & 1.35 & 0.9\% & 2.5 \\
CAA & PEN/Hierarchical & 70 & 2.5\% & \textbf{1.4} & \textbf{0.9}\% & \textbf{2.6} \\
ECAA & Hierarchical 80 & 200 & 1\% & \textbf{1.65} & \textbf{0.3}\% & \textbf{3.5} \\
ECAA & SVN & 1 & 0.5\%& 0.4 & \textbf{0.8}\% & 0.6 \\
\bottomrule
\end{tabular}
\end{adjustbox}
}
\end{table}

We obtained optimistic results - especially for the downside loss (see \ref{loss2}) which is more appropriate in this framework. We evaluated the performance using four well established portfolio risk measures: the return of the portfolio, sharpe ratio  is  the  amount of return an investor receives per unit of risk, the maximum drawdown is  the maximum observed loss from a peak to a trough of a portfolio, before a new peak is attained and calmar ratio  measures the risk-adjusted performance of a portfolio by comparing the return to the   maximum drawdown.

 The distribution of traders' returns is close to symmetric and the mean is approximately zero. Performances of CAA are on the whole comparable with those of the MEAN clustering decision rule for the clusters constructed with the  SVN - infomap method. However the results using the hierarchical clustering are significantly better across all risk measures. The best performing cutoff for the distance metric is around $70\%$. On the other hand, the results for the PEN clustering decision rule are comparable for the return on investment but for other metrics we  noticed significantly better results for both clustering techniques.
 Figures  \ref{fig:outsamplepen}  and \ref{fig:outsamplemean}  show the comparison among all results for a return scaling factor up to 400.

\begin{figure}[htbp]
\floatconts
 {fig:outsamplepen}
  {\caption{Comparison of results among all four considered measures of risk in the out of  sample scenario where the  CAA learner prediction is the experts predictions divided by the cardinality of  each cluster. The return to  maximum drawdown ratio, sharpe ratio, 1 +return and maximum drawdowm  are shown  for different return scaling factors. The green,blue and pink dotted line denote the equal weights portfolio, AA and CAA for SVN- infomap performances. Other  curves represent CAA using clusters done with hierarchical clustering  with different thresholds. }}
  {  \includegraphics[width=.95\linewidth]{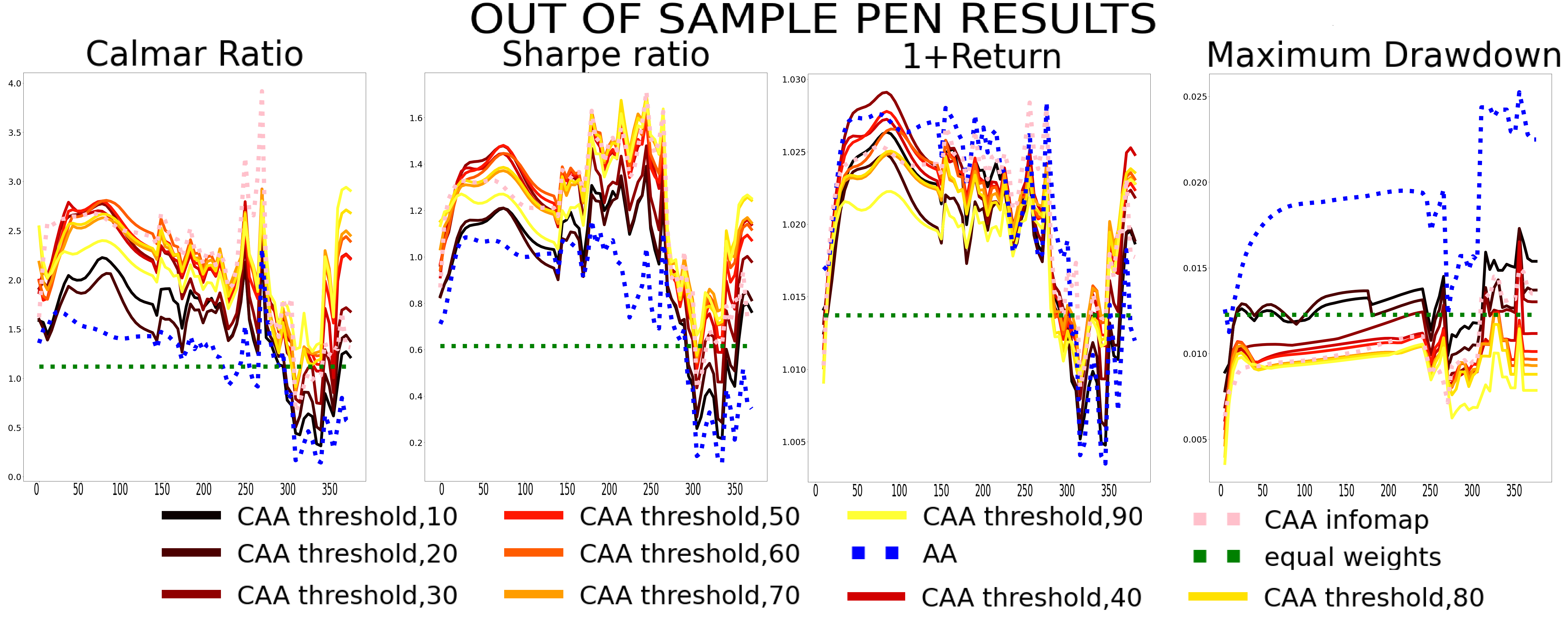}}
\end{figure}

\begin{figure}[htbp]
\floatconts
 {fig:outsamplemean}
  {\caption{Comparison of results among all four considered measures of risk in the out of  sample scenario where the  CAA learner prediction is the  mean  of experts prediction for  each cluster. The return to  maximum drawdown ratio, sharpe ratio, 1 +return and maximum drawdowm  are shown  for different return scaling factors. The green,blue and pink dotted line denote the equal weights portfolio, AA and CAA for SVN- infomap performances. Other  curves represent CAA using clusters done with hierarchical clustering  with different thresholds. }}
  {   \includegraphics[width=.95\linewidth]{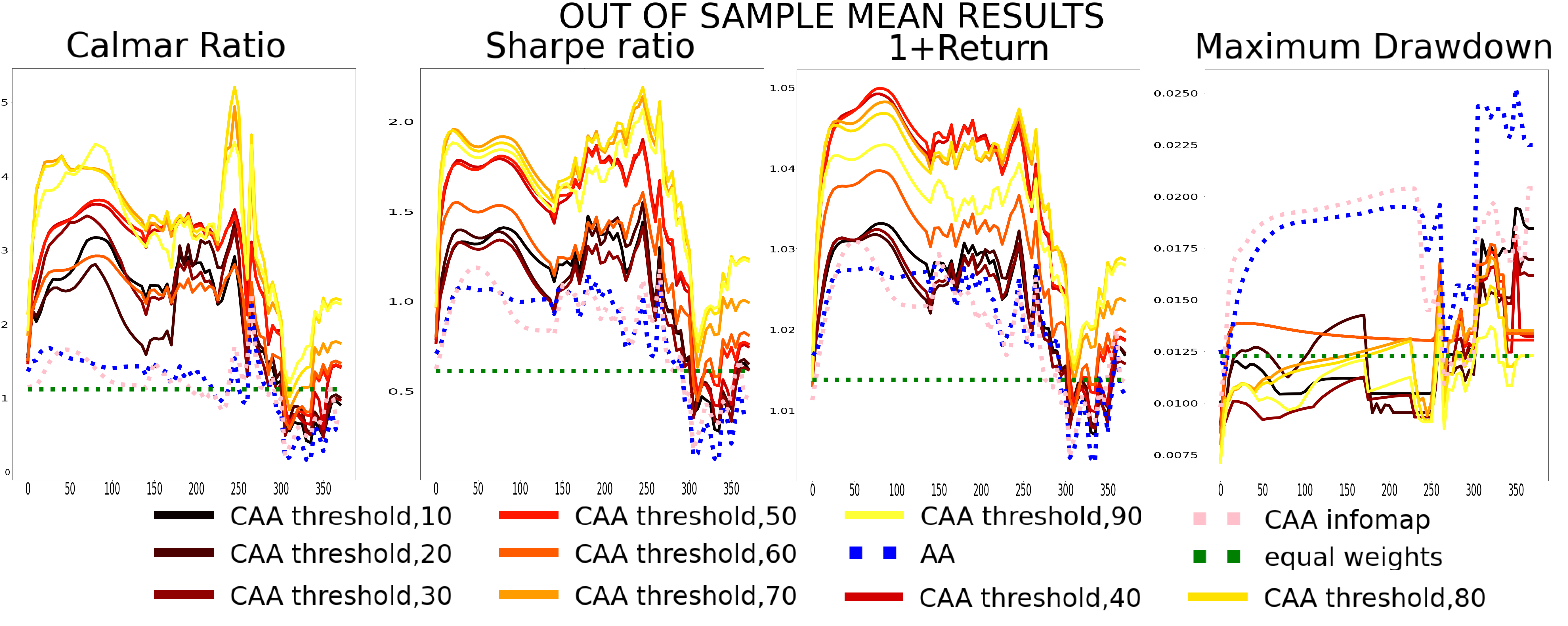}}
\end{figure}

For ECAA we consider the scenario of treating clusters as meta-experts. Using the alluvial chart we  can readily identify the flow of clusters over time since without it we could not identify clusters at different time epochs since they are unlabeled. Overall performance of the ECAA using SVN-infomap clusters is poor, manifesting lowest return, Sharpe Ratio and Calma Ratio  . However for  hierarchical clustering all other risk measures are significantly better than the standard AA besides the return (see Figure \ref{fig:outsampleaasquared}). Moreover, ECAA has smoother PnL as seen by much smaller drawdown than CAA, AA and the banchmark.

\begin{figure}[htbp]
\floatconts
 {fig:outsampleaasquared}
  { \caption{Comparison of results among all four considered measures of risk in the out of  sample scenario where the  ECAA learner prediction is the  mean  of experts prediction for  each cluster. The return to  maximum drawdown ratio, sharpe ratio, 1 +return and maximum drawdowm  are shown  for different return scaling factors. The green,blue and pink dotted line denote the equal weights portfolio, AA and ECAA for SVN- infomap performances. Other  curves represent ECAA using clusters done with hierarchical clustering  with different thresholds. }}
  {   \includegraphics[width=.95\linewidth]{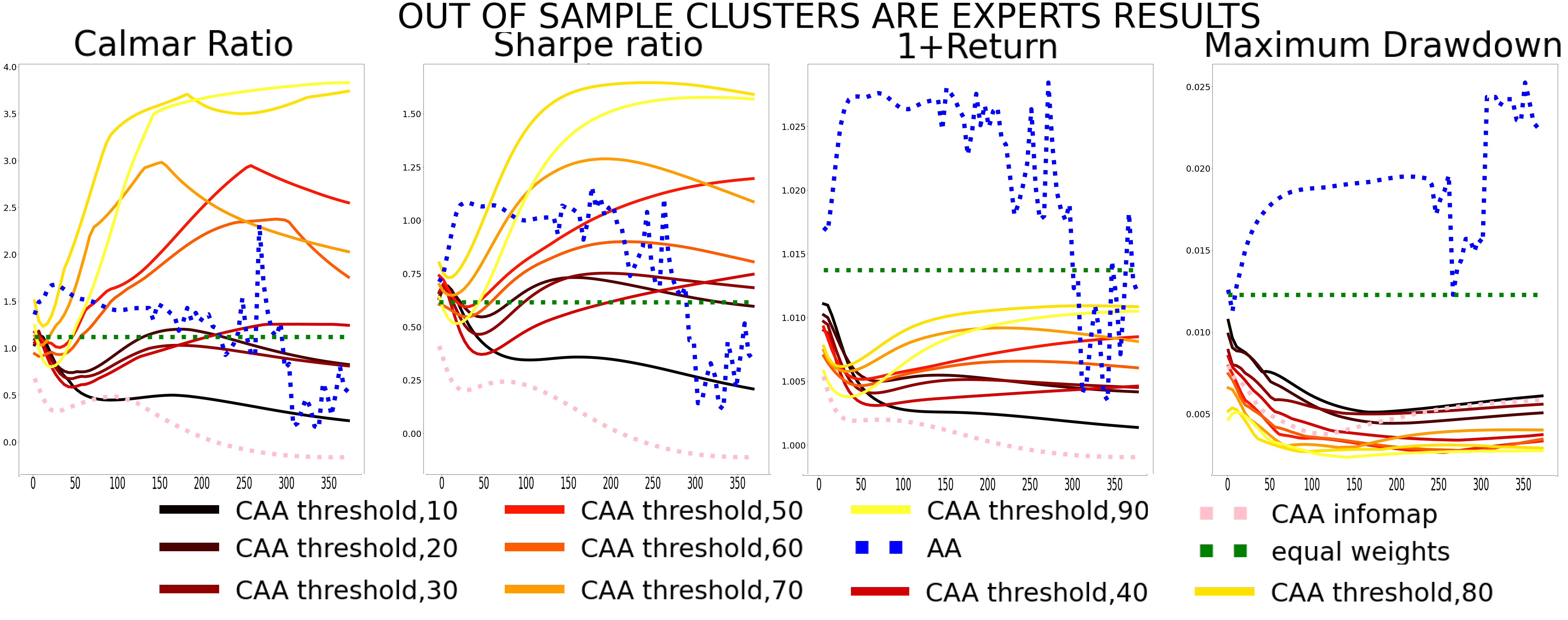}}
\end{figure}

\begin{figure}[htbp]
  \floatconts
    {fig:both3}
    { }
    {\begin{minipage}[b]{0.45\linewidth}
      \centering
      \includegraphics[width=0.95\textwidth]{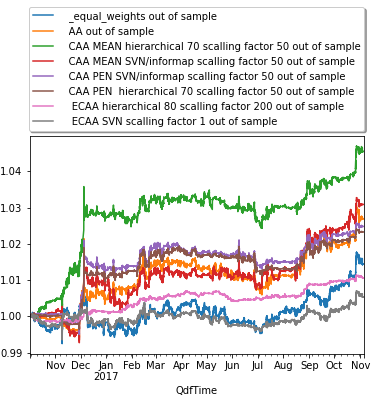}
      \captionsetup{justification=centering}
       \caption{ Comparison of  returns for equal weight portfolio , AA and good alternatives for CAA and ECAA.}
     \label{fig:inoutsample}  
    \end{minipage}
    \hfill
    \begin{minipage}[b]{0.45\linewidth}
      \centering
        \includegraphics[width=0.95\textwidth]{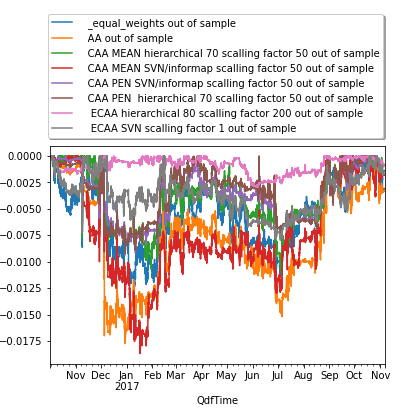}
      \captionsetup{justification=centering}
      \caption{ Comparison of  relative drawdowns  for equal weight portfolio , AA and good alternatives for CAA and ECAA.  }
      \label{fig:inoutsampledd} 
    \end{minipage}}
\end{figure}

Table \ref{Tbl:summary} summarises the experimental results for near optimal variations of all algorithms. Figures \ref{fig:inoutsample} and \ref{fig:inoutsampledd} show their evolution of returns and drawdowns throughout time. It is worth mentioning that when the scaling factor gets bigger (larger than $100$) more and more traders go bankrupt because of the nature of the loss (\ref{loss2}). Moreover, the algorithm could suddenly stop investing when the scaling factor gets too big therefore one must be cautious when interpreting the results.

\section{Conclusion}
 In this paper our findings confirm that clustering of traders' investments can be  described by Ewens distribution. The temporal clustering distribution depends on many  parameters and market conditions however its clustering could be leveraged to make better investment decisions. We adjusted the aggregating algorithm  with sleeping experts to test the latter hypothesis using two clustering techniques, namely SVN-infomap and hierarchical clustering. In this framework the latter approach gives better results and gives more meaningful clusters since is  based on correlations of the investors' net positions and not on their trading synchronicity.
 In particular we compared CAA (used aggregated traders' decisions per cluster to calculate the investment prediction) and ECAA (clusters played the role of experts) with AA and the equally weighted portfolio strategy. Our introduced modifications to the AA indicate clear performance benefits in our experimental results in terms of  four well established portfolio risk measures: return, Sharpe ratio, maximal drawdown and Calmar ratio.

\comment{ 
\begin{figure}[htbp]
\floatconts
  {fig:qq}
  {\caption{QQ plot for change of net position. Left subplot is after Yeo-Johnson transformation.The right is after normalazing transformation.  }}
  {\includegraphics[width=12cm]{images/quantiles.png}}
\end{figure}

\begin{table}[htbp]
\floatconts
  {tab:test}
  {\caption{Statistical coverage test results. The percentage score is  the average success rate of passed tests across all considered  significance levels (19 levels from 0.05 to 0.95). }}%
  {%
\footnotesize
\begin{tabular}{lllll}
\toprule
                Predictor &        Type &  UC &   CC \\
\midrule
    Gradient  Boosting &          CP &  79\% &     0\% \\
    Gradient Boosting &         NCP &  21\% &      5\% \\
    Gradient Boosting &    Q &  68\% &       0\% \\
 K-Nearest Neighbours &          CP &     0\% &  0\% \\
 K-Nearest Neighbours &         NCP &      0\% &  0\% \\
 K-Nearest Neighbours &    Q &   5\% &     0\% \\
                LSTM &          CP &  74\% &    0\% \\
                LSTM &         NCP &  11\% &     0\% \\
                LSTM &    Q &   0\% &   0\% \\
     Linear Regression &          CP &  11\% &     0\% \\
     Linear Regression &         NCP &  11\% &     0\% \\
     Linear Regression &    Q &  11\% &    0\% \\
                  MA &  Benchmark &   0\% &     0\% \\
        RandomForest &          CP &  11\% &     0\% \\
        RandomForest &         NCP &  21\% &     0\% \\
        RandomForest &    Q &  11\% &      0\% \\
\bottomrule
\end{tabular}
  }
\end{table}

}

\acks{The authors acknowledge the support of Algorithmic Laboratories Ltd (AlgoLabs) and their their parent company Equiti Group in establishing and developing this research. Special thanks go to Xudong Li, Tzyy Tong and Samuel Manoharan for setting up the servers necessary to run our experiments. Further thanks go to Simon Tavar\'e for useful insights. }


\appendix

\section*{Appendix: Clusterised AA bound  \label{CAAbound_apendix}}

In this section, we will discuss when it is beneficial to run AA on (equally weighted) cluster experts rather than the original experts and connect this with our intuition about the performance of traders. The analysis will be done on an artificial example but the conclusion is instructive.

Suppose that we have $m$ identical experts in a pool of $N$. One may want to collate them into
one; there is no need though as this is done by the AA automatically. The behaviour of the
AA would be the same as if one expert with the combined weight is present in
the pool. Assuming the uniform distribution on  $N$ original experts, the weight
of the combined expert will be $m/N$ and the loss bound for the duplicated experts
$E_i$ (assuming the mixable case $C = 1$) turns into
$$\Loss_T(L) \leq \Loss_T(E_i)+\frac{1}{\eta}\ln\frac{N}{m}.$$
This is a stronger bound and if the performance of the expert is actually good, it leads to lower $\Loss_T(L)$. However, if duplicate experts perform badly, they create a problem: increasing $N$ worsens the bound for good experts.

Suppose that we have $M$ clusters of experts  of cardinalities $c_1,..,c_M$. Let all experts in each cluster be identical and suffer the same cumulative loss. Applying AA to cluster meta experts (with equal initial weights) will give us the loss bound
$U_{-}$ and applying AA to the original experts will give us the loss bound $U_{*}$: 
\begin{eqnarray}
U_{-}&= & \min_{i=1,2,\ldots,M} \Big\{\Loss_T(E_{C_i})+ \frac{1}{\eta}\ln M\Big\}=\Loss_T(E_*)+ \frac{1}{\eta}\ln M,\nonumber
\\
{U_*}&= & \min_{i=1,2,\ldots,M} \Big\{\Loss_T(E_{C_i})+ \frac{1}{\eta}\ln\frac{N}{c_i}\Big\}=\Loss_T(E_{C_{i_0}})+ \frac{1}{\eta}\ln\frac{N}{c_{i_0}},\nonumber
 \end{eqnarray}
where $E_{C_i}$ is an expert from cluster $i$, $E_*$ is the best expert overall, and $i_0$ is the number of the cluster where the minimum in $U_*$ is achieved.
 
We get that
 \begin{equation}
 \hspace{1.23in} U_{-} \leq U_* \iff c_{i_0}\leq \frac{N}{M} e^{\eta[\Loss_T(E_{C_{i_0}})-\Loss_T(E_{*})]},    \label{eq:conditionAA}
\end{equation}
where $\Loss_T(E_{C_{i_0}})-\Loss_T(E_{*})\ge 0$. This means that the bound with cluster meta experts is better when there are no good experts in large clusters.

As the practice of trading shows, good trades are usually few and make a minority, which is one of the justification for the cluster AA. Cluster AA gives an advantage to smaller clusters.

 

\end{document}